\documentclass[aps,superscriptaddress,twocolumn]{revtex4-1} 
\usepackage{latexsym}
\usepackage{graphicx}
\usepackage[margin=0.5in]{geometry}
\usepackage{amsmath}
\usepackage{amssymb}
\usepackage{verbatim}
\usepackage{braket}
\usepackage{upgreek}
\usepackage{float}
\usepackage{breqn}
\usepackage{subfig}
\usepackage{setspace}
\usepackage{color}
\usepackage[compact]{titlesec} 
\titlespacing{\section}{0pt}{0pt}{0pt} 
\AtBeginDocument{
	\setlength\abovedisplayskip{0pt}
	\setlength\belowdisplayskip{0pt}}
\usepackage{hyperref}
\date{\today}
\makeatletter
\let\cat@comma@active\@empty
\makeatother
\begin{document}
	\title{Topological Quantum Criticality in non-Hermitian Kitaev chain with Longer Range Interaction}
	
	\author{S Rahul}
		\affiliation{Department of Theoretical Sciences, Poornaprajna Institute of Scientific Research, Bidalur, Bengaluru -562164, India.}
	\affiliation{Manipal Academy of Higher Education, Madhava Nagar, Manipal - 576104, India.}
	\author{Sujit Sarkar}
	\affiliation{Department of Theoretical Sciences, Poornaprajna Institute of Scientific Research, Bidalur, Bengaluru -562164, India.}
	
	\begin{abstract}
	\noindent An Attempt is made to study the non-Hermitian effect on the topological quantum criticality and also in the physics of Majroana zero mode (MZMs).		
In this work, the effects and modifications done by the non-Hermitian factor $\gamma$ on the topological phases, criticality and also in the MZMs is studied. We use the zero mode solutions (ZMS) to construct the phase diagram. We find a correspondence between Hermitian and non-Hermitian model Hamiltonian. The MZMs appear at criticality and has a stability dependence on the new passage created because of the introduction of non-Hermiticity. The multicritical points are also studied to understand their nature under the influence of non-Hemiticity. We also study the effect of non-Hermiticity on the topological phases.

		
	\end{abstract}
	\maketitle
	\section{Introduction}
	The study of Non-Hermitian systems have gained an immense importance in the recent times when it entered the area of topological systems. Few of the important aspects which makes non-Hermitian systems to standout are complex energy dispersion, exceptional points and modified bulk boundary correspondence \cite{shen2018topological,bergholtz2019exceptional,kunst2018biorthogonal,kunst2018biorthogonal}. Non-Hermitian system offers a look into a more realistic and intereactive setup, in other words, the non-Hermiticity is the interaction of environment with the system or vice versa. $PT$ symmetric Hamiltonians are the subclass of non-Hermitian systems, where, these Hamiltonians possess real eigenvalues in the $PT$ unbroken phase. These $PT$ symmetric Hamiltonians are widely used in the field of optics because, controlled dissipation can be achieved in these systems \cite{ruter2010observation,peng2014parity,xiao2017observation,ashida2020non,cai2021boundary,el2019dawn,el2018non}.\\ 
	 There have been studies on the non-Hermitian Kitaev system hosting MZMs \cite{wang2015spontaneous,zeng2016non,li2018topological}. Topological properties and topological invariant number of non-Hermitian system \cite{lieu2018topological,zhu2014pt,he2020non,yao2018edge,yin2018geometrical,navarro2021geometrical} is studied where they show stable topological phases and fractional topological invariant number. 
	 The extended Kitaev model (Hermitian) hosts more than one topological phases which is in a way responsible for the appearance of MZMs at criticality. 
	 It consists of non-high symmetry critical line where the the excitation dispersion is quadratic in nature whereas in all other critical lines that it possess has linear dispersion.
	 The extended Kitaev model also possesses multicritical point in which one of them is responsible for the topological phase transition on the critical line to occur. With these important points, we intend to study the effects of non-hermiticity on the non-Hermitian counterpart of the extended Kitaev model. The non-Hermitian counterpart of the extended Kitaev model has $i\gamma$ term added to its on-site chemical potential $\mu$.\\
	 This paper is organized as follows, in section.\ref{one} the model Hamiltonian and its complex energy spectrum are discussed.\@ In section.\ref{comp}, the ZMS, critical line computation and topological phase transitions across gapped phases is discussed.\@ Using ZMS the phase diagram is constructed topological phases and critical lines are characterized.\@ In subsection.\ref{topo2} topological phase transitions are analyzed using parametric curves which shows the point of transition from one phase to the other.\@ In subsection.\ref{topo3}, MZMs at criticality is studied and one to one comparison with its Hermitian counterpart is derived.\@ The displacement of multicritical point in the non-Hermitian system when compared to its Hermitian counterpart has been extensively investigated in the subsection.\ref{topo1}.\@ The physics of the multicritical point opens up a new way of thinking about the stability problem in the topological systems.\@ The effect of non-Hermiticity on the topological phases and criticalities are studied with increasing the values of non-Hermitian factor.\@ The distinct behavior of critical lines has been investigated in the subsection.\ref{topo}.   
   \section{Model Hamiltonian}\label{one}
 We consider an extended Kitaev chain in presence of non-Hermiticity given as,
\begin{dmath}
	H = - \lambda_1 \sum_{i=1}^{N-1} (c_{i}^{\dagger}c_{i+1} + c_{i}^{\dagger}c_{i+1}^{\dagger} + h.c) - \lambda_2 \sum_{i=1}^{N-1} ( c_{i-1}^{\dagger}c_{i+1} +  c_{i+1} c_{i-1} + h.c) -(\mu+i \gamma) \sum_{i=1}^{N} (1 - 2 c_{i}^{\dagger}c_{i}). 
	\label{jw1} 
\end{dmath} 
Where $\lambda_1$, $\lambda_2$, $\mu$ and $\gamma$ corresponds to nearest, next nearest neighbor coupling, chemical potential and non-Hermitian factor respectively.\\
The Bogoliubov de-Gennes (BdG) form of the Hamiltonian given by,
\begin{dmath}
H_k =  \chi_{z} (k) \sigma_z - \chi_{y} (k) \sigma_y = \left(\begin{matrix}
\chi_{z} (k) && i\chi_{y} (k)\\
-i\chi_{y} (k) && -\chi_{z} (k)\\
\end{matrix} \right),
\label{APS}
\end{dmath}
where $ \chi_{z} (k) = -2 \lambda_1 \cos k - 2 \lambda_2 \cos 2k + 2(\mu+ i \gamma),$ and $ \chi_{y} (k) = 2 \lambda_1 \sin k + 2 \lambda_2 \sin 2k.$ 
Energy dispersion relation is,
\begin{equation}
E_k=\pm \sqrt{(\chi_{z} (k))^2 + (\chi_{y} (k))^2}.
\label{es}
\end{equation}    
The energy eigenvalues are complex throughout the parameter space and hence the model Hamiltonian Eq.\ref{APS} does not obey $PT$ symmetry.\\ It can be verified using the $PT$ symmetry operator $\hat{K} \sigma_z$ acting on the Hamiltonian $H$, i.e., $\hat{K} \sigma_z \hat{H_k} \hat{K^{-1}} \sigma_z \ne \hat{H_k} $. Calculating the critical lines becomes tedious for the model Hamiltonian and hence ZMS are used to characterize the topological phases and criticalities. 
\section{Results and Discussions}\label{comp}
We derive the ZMS from the model Hamiltonian which is relegated in the appendix.\\
 The ZMS are given as,
\begin{equation}
X_{\pm} = \frac{-\lambda_1 \pm \sqrt{\lambda_1^2 + 4 \lambda_2 (\mu+i \gamma)}}{2 \lambda_2}. 
\label{12}
\end{equation}
These ZMS (Eq.\ref{12}) are nothing but the zeros of a model Hamiltonian that can be written in the form of a complex function \cite{rahul2021majorana,verresen2018topology}. Zeros lying inside, outside and on the unit circle indicates topological, non-topological phase and criticality respectively. In this study, the ZMS are studied as a function of $\lambda_1$ to understand the topological phases and phase boundaries. The absolute value of ZMS is considered becaused the roots are complex in nature. 
\begin{figure}[H]
	\centering
	\includegraphics[scale=0.35]{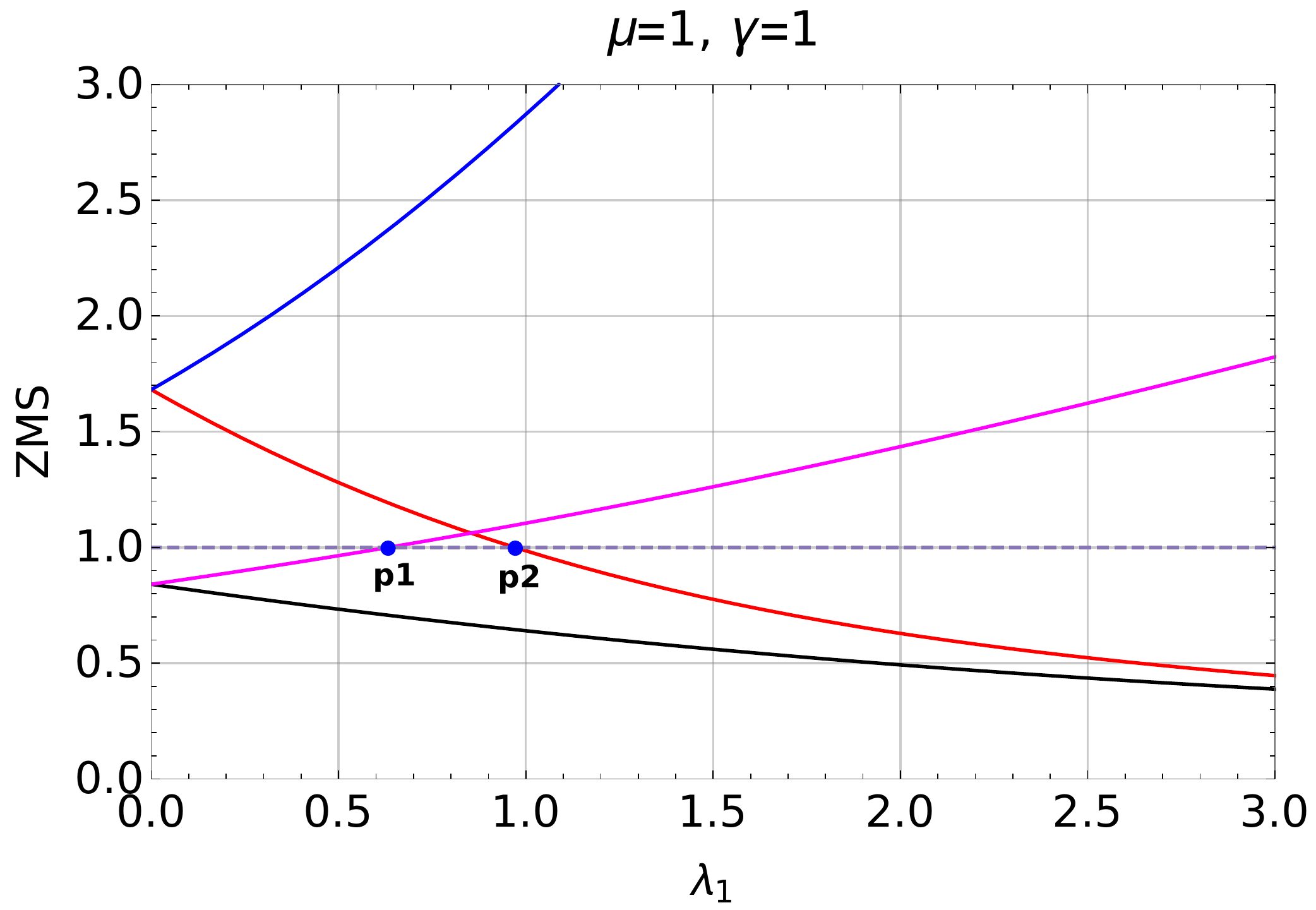}
	\caption{Zero mode solutions plotted with respect to the parameter $\lambda_1$ shows both $w = 0$ to $w = 1$ $(\lambda_2 = 0.5)$ and $w=1$ to $w=2$ $(\lambda_2 = 2.0)$ topological phase transitions. Two blue dots (p1 and p2) represent the transition points.}
	\label{f2}
\end{figure}
In the Fig.\ref{f2}, red and blue curves are the roots representing $w=0$ to $w=1$ transition and magenta and black curves are the roots representing $w=1$ to $w=2$ transition.  
To specify the transition point using ZMS, parallel to the x-axis, a reference line (unit line), $y=1$ is drawn which acts in a same way as that of the unit circle drawn to analyze the zeros of a complex function. The point of the intersection between one of the ZMS and the unit line marks the transition point which is represented in blue dots ($p1$ and $p2$) in the Fig.\ref{f2}. Value of the roots greater than 1 corresponds to the non-topological phase whereas less than 1 corresponds to topological phase. Transition points $p1$ and $p2$ marks the transitions between $w=2$ to $w=1$ and $w=0$ to $w=1$ respectively with fixed positive value of $\lambda_2$ ($\lambda_2$ = 0.5, 2.0). Keeping track of these transition points via the ZMS method provides the phase diagram of the model Hamiltonian. 
\begin{figure}[H]
	\centering
	\includegraphics[scale=0.23]{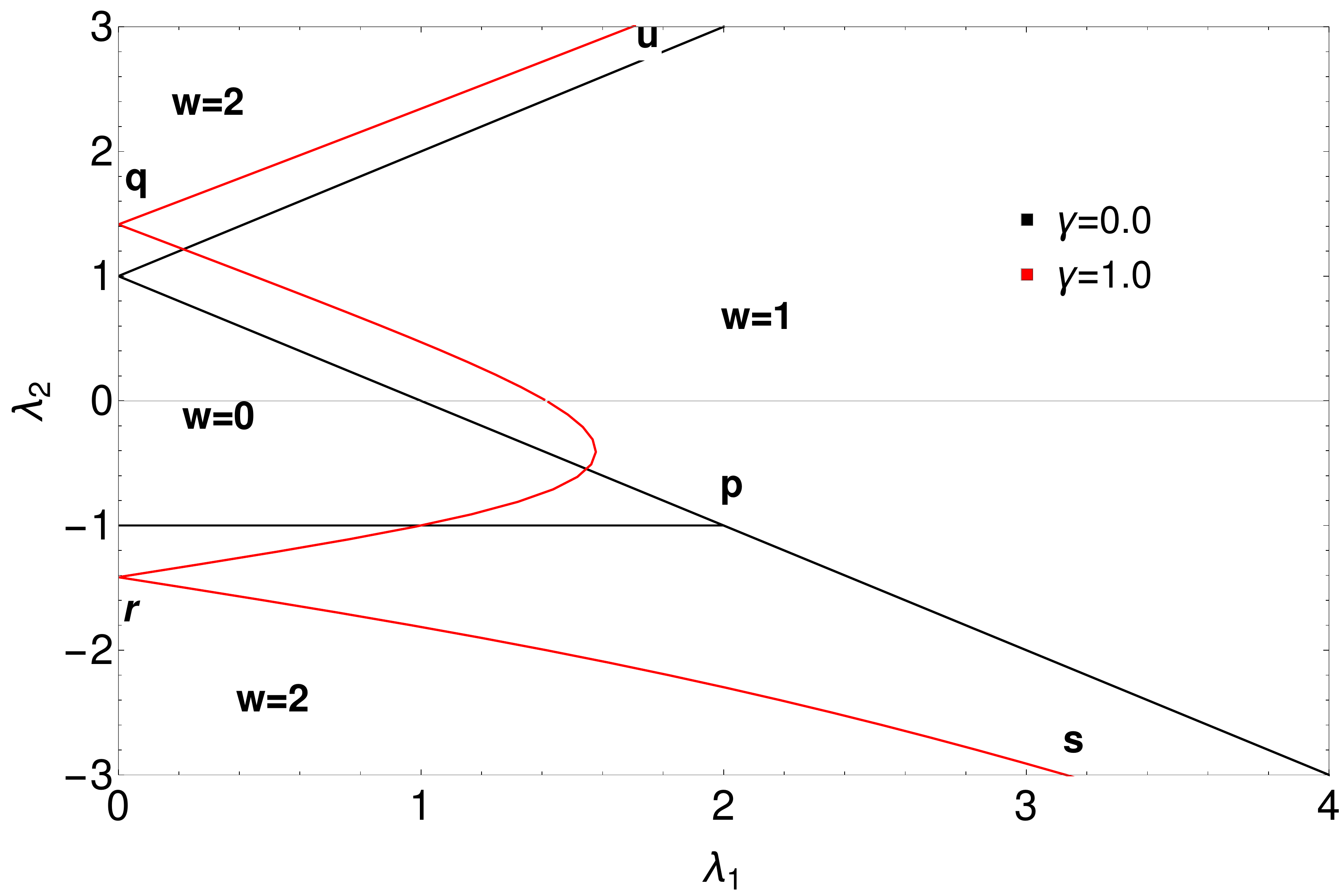}
	\caption{Phase diagram of the model Hamiltonian in presence of $\gamma$ $(=1.0)$. Black lines corresponds to the phase diagram for $\gamma=0.0$ (Hermitian case).}
	\label{pd}
\end{figure}
In the positive $\lambda_2$ region of phase diagram Fig.\ref{pd}, the effect of $\gamma$ on these critical lines is just a parallel shift. In the negative $\lambda_2$ region, numerically computed critical points shows that the critical line follows the curved path (q to r) as shown in the Fig.\ref{pd}. The critical line (r to s) also follows a curved path.
\subsection{Parametric curve analysis in the negative $\lambda_2$ region}\label{topo2}
\begin{figure}[H]
	\centering
	\includegraphics[width=8cm,height=5cm]{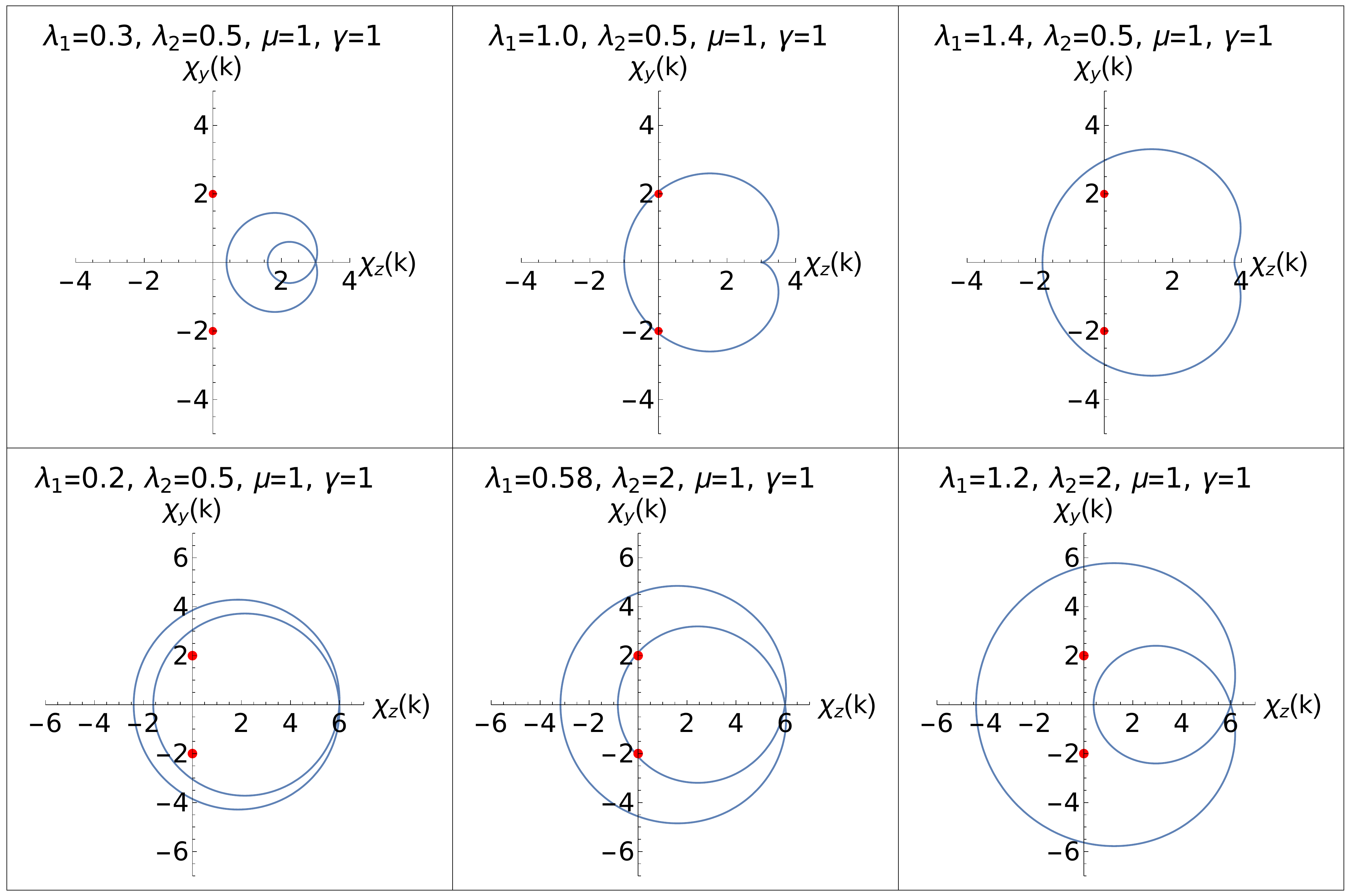}
	\caption{Parametric plots representing the topological phase transitions in the positive $\lambda_2$ region. Upper and lower panel corresponds to $w=0$ to $w=1$ and $w=2$ to $w=1$ respectively. Middle plot in both the panels represents the parametric curves at critical points.}
	\label{para1}
\end{figure}

\begin{figure}[H]
	\centering
	\includegraphics[width=8.5cm,height=4.6cm]{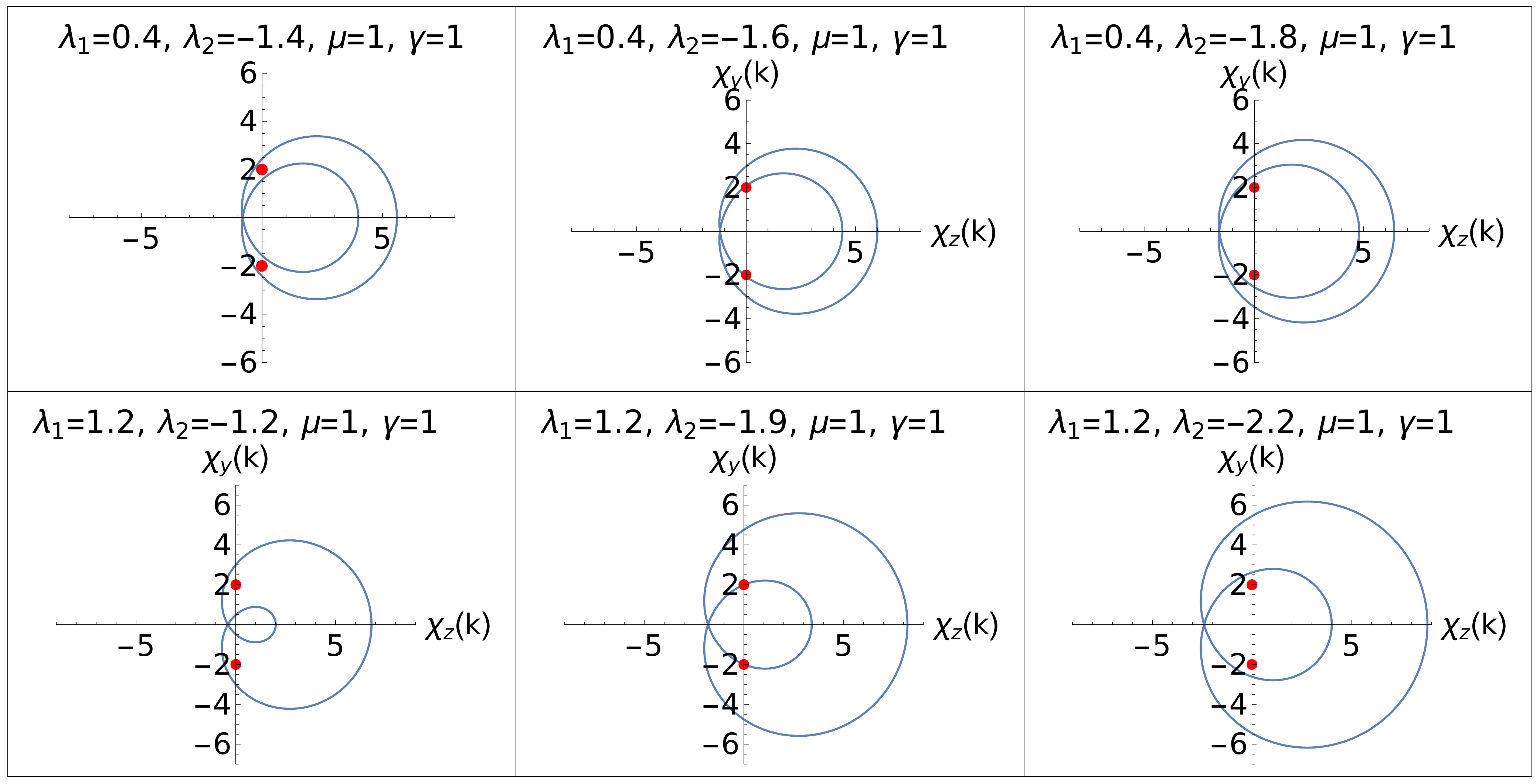}
	\caption{Parametric plots representing the topological phase transitions in the negative $\lambda_2$ region. Upper and lower panel corresponds to $w=1$ to $2$ for different values of $\lambda_1$ respectively. Middle plot in both the panels represents the parametric curves at critical points.}
	\label{para3}
\end{figure}

In the Hermitian systems, the topological phase transition is marked when the parametric curve touches the origin but 
in the case of non-Hermitian systems, the transition is marked when the parametric curve touches the Eps. These Eps are special kind of degenerate points at which the excitation energy becomes zero, i.e., $h_{x}^2 + h_{y}^2 = 0 $. The positions of EPs are given as, 
$$h_{xr} = - h_{yi},\;\;\; and \;\;\; h_{yr} = h_{xi}$$\\
or 
$$h_{xr} = h_{yi},\;\;\; and \;\;\; h_{yr} = -h_{xi}.$$
In the Fig.\ref{para1} we study the parametric curves for $w=0$ to $1$ and $w=2$ to $w=1$ topological phase transition. Two red spots on the y-axis at 2,-2 are the positions of exceptional points where exceptional points are the critical points through which the transition occurs. Just as in the case of Hermitian systems, parametric curve depicts the transition when it touches the origin, similarly in the non-Hermitian systems, the transition is marked when the parametric curves touch the exceptional points which can be seen in the Fig.\ref{para1}.
 Fig.\ref{para1} only shows the topological phase transitions between the topological phases in the positive $\lambda_2$ region. In the upper panel of the Fig.\ref{para1}, gapped phase $w=0$ (left plot), transition point (middle plot) and gapped phase $w=1$ (right plot) are shown. 
\begin{figure}[H]
	\centering
	\includegraphics[width=9cm,height=3.8cm]{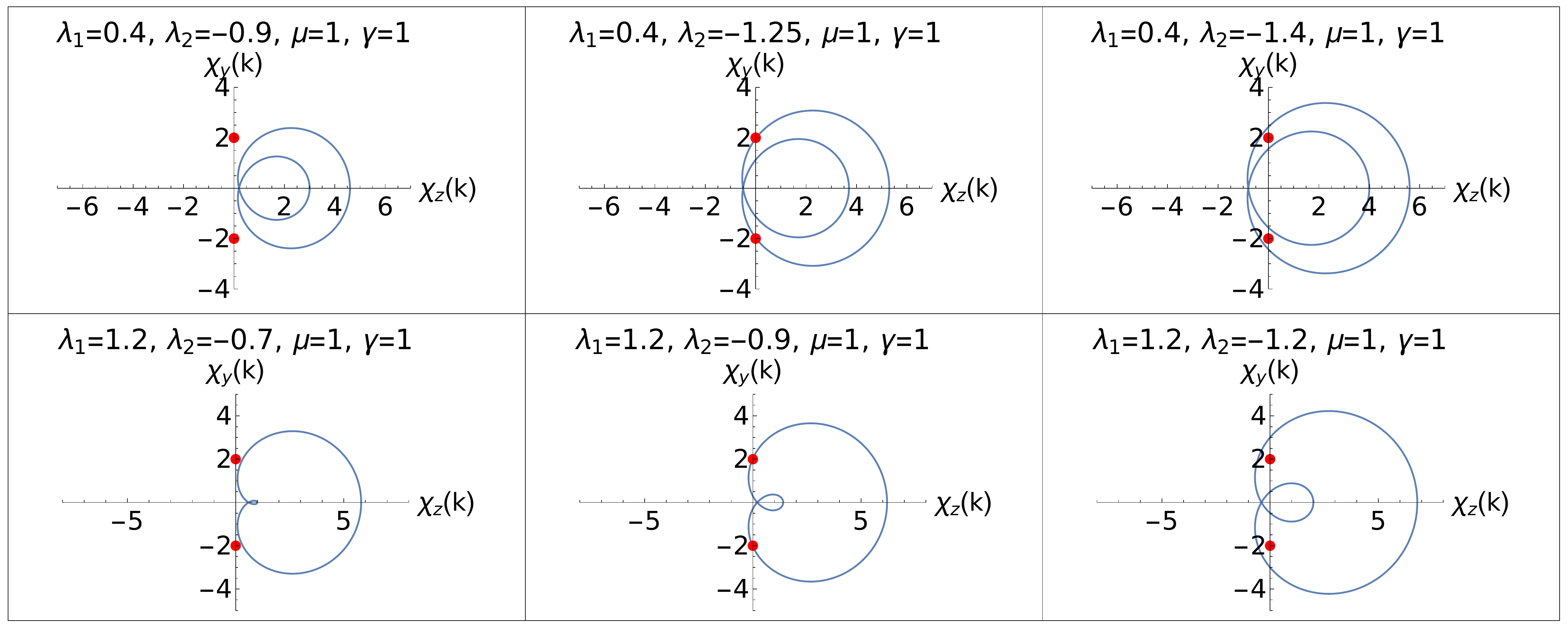}
	\caption{Parametric plots representing the topological phase transitions in the negative $\lambda_2$ region. Upper and lower panel corresponds to $0$ to $1$ for different values of $\lambda_1$ respectively. Middle plot in both the panels represents the parametric curves at critical points.}
	\label{para2}
\end{figure} 
Similarly in the lower panel, transition from $w=2$ to $w=1$ is depicted. From the numerical calculations of the transition points as described in the section.\ref{comp}, it is seen that the critical lines in the negative $\lambda_2$ region behave in a significantly different manner when compared to the positive $\lambda_2$ region. In order to understand the topological phase transitions in the negative $\lambda_2$ region, the parametric curves are plotted to depict the transition between gapped phases as shown in the Fig.\ref{para2} and Fig.\ref{para3}. 
The deflection of the critical line "qr" as shown in the phase diagram Fig.\ref{pd} is novel behavior observed in the non-Hermitian system. Fig.\ref{para2} shows the transition between $w=0$ to $w=1$. Similarly, parametric plot representing $w=1$ to $2$ for two different values of $\lambda_1$ is also shown in the Fig.\ref{para3}.
Fig.\ref{para3} presents the topological phase transition between $w=1$ and $2$ for two different values of $\lambda_1$. Upper panel is for $\lambda_1 = 0.4$ where left plot corresponds to $w=1$ phase, middle plot corresponds the transition and the right plot corresponds to topological phase $w=2$.\\ 
Parametric curves showed the gap closing at the topological phase transition points for two different values of $\lambda_1$. Deviated critical lines shown in the phase diagram Fig.\ref{pd} is suppprted by the parametric curves. Due to this deviation of critical line, the multicritical point is dispaced and as a result, the nature of the multiciritcal point is also changed which will be explored in the further section. 
\subsection{Majorana zero modes at Criticality}\label{topo3}
  For the non-Hermitian model Hamiltonian considered, MZMs are present in their respetive topological phases just as in the case of Hermitian systems. Authors of reference.\cite{rahul2021majorana} have studied the characterization and MZMs at criticality and similarly in the non-Hermitian systems, we observe the MZMs at criticality. 
  The MZMs at criticality is studied using ZMS described in the section.\ref{comp}.
  Critical lines, "qu" and "rs",shown in the phase diagram Fig.\ref{pd} hosts the MZMs becuase these critical lines are the phase boundaries between topological phases $w=2$ and $w=1$. 
  \begin{figure}[H]
  	\centering
  	\includegraphics[scale=0.4]{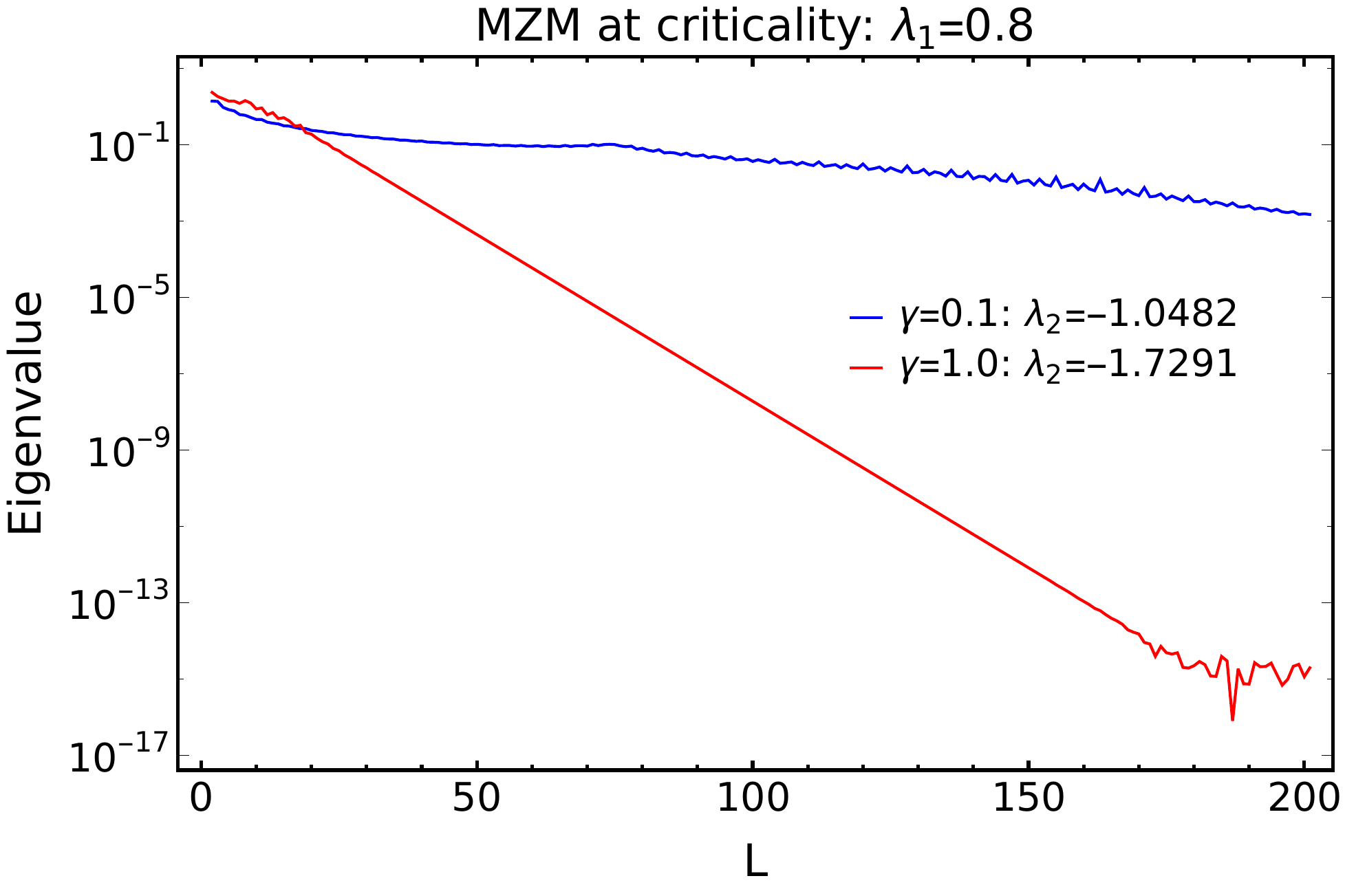}
  	\caption{System size plot of the zero mode eigenvalues for different values of $\gamma$.}
  	\label{mzmcrit}
  \end{figure}
  With the introduction of $\gamma$, the multicritical point gets displaced and a passage is opened between the critical lines "qr" and "rs" which can be seen from the Fig.\ref{pd2}. This passage acquires  $w=1$ topological phase after characterizing and hence the critical line "rs" hosts the MZMs at criticality because after the introduction of $\gamma$ it is the phase boundary between the topological phases $w=1$ and $w=2$. 
  \begin{figure}[H]
  	\centering
  	\includegraphics[scale=0.25]{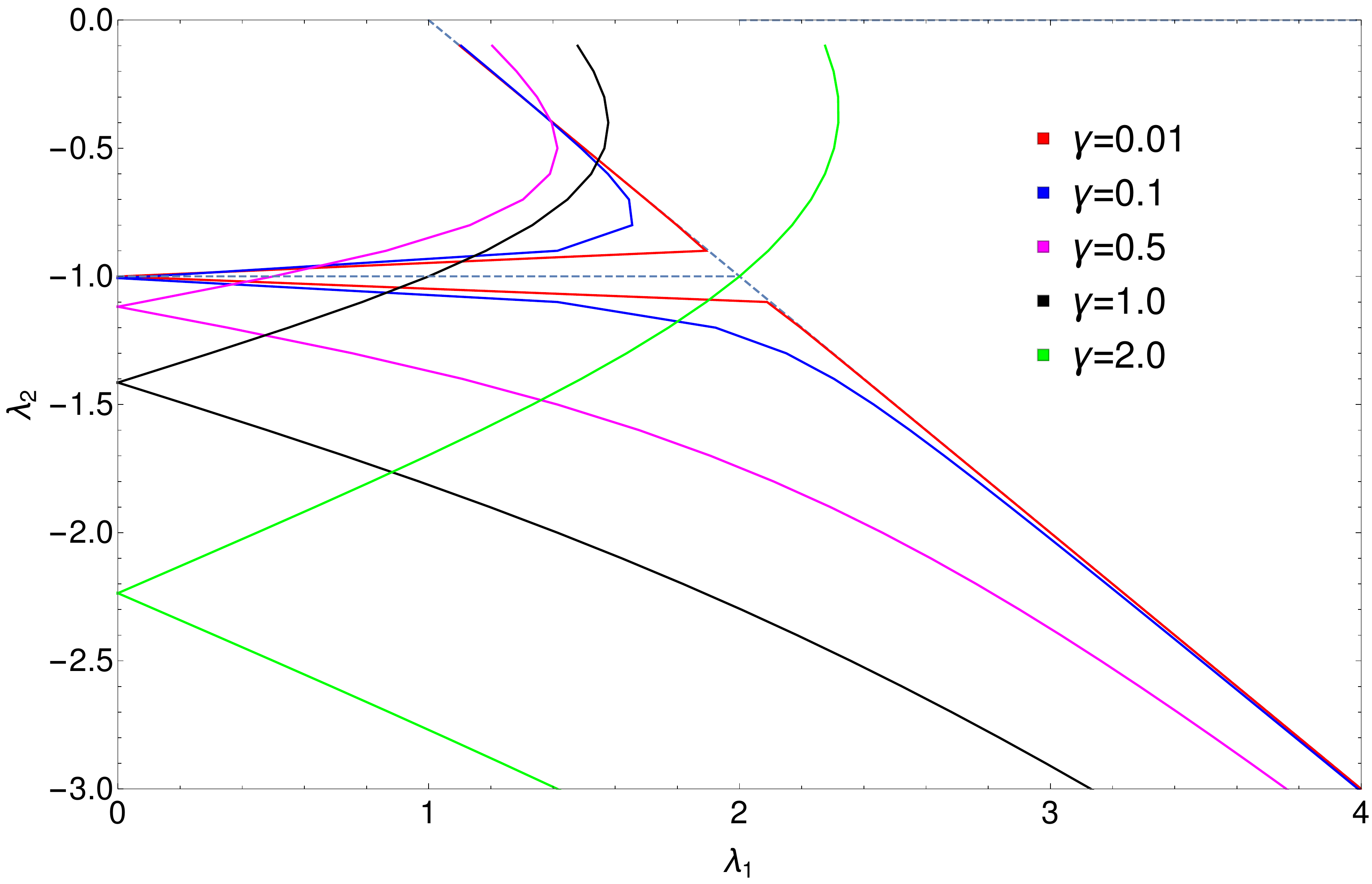}
  	\caption{Negative region of $\lambda_2$ plotted for different values of $\gamma$. All the intersecting points on the negative $\lambda_2$ axis are the positions of multicritical point for different values of $\gamma$.}
  	\label{pass}
  \end{figure}
  In the positive $\lambda_2$ region, the critical line "qu" also hosts the MZMs at criticality since it is the phase boundary between $w=1$ and $w=2$. For smaller value of $\gamma$, the passage is small in the negative $\lambda_2$ region and as the $\gamma$ tends to zero, the passage (refer Fig.\ref{pass}) disappears and a single critical line separating $w=0$ and $w=2$ topological phases appear which is nothing but the Hermitian case. Hence for smaller values of $\gamma$ $(\gamma=0.1)$, the $w=1$ topological phase (passage) is not stable and hence the MZMs at criticality is also less stable. For $\gamma = 1$, the passage becomes wider and more stable hence the MZMs at criticality are also stable. The stability of the MZMs at criticality is presented in the Fig.\ref{mzmcrit}.
  In the Fig.\ref{mzmcrit}, Both the zero mode eigenvalues show exponential decay with respect to the system size. For $\gamma = 0.1$, zero mode eigenvalue (blue) decay with respect to the system size is slow compared to the decay rate of the eigenvalue (red) for $\gamma=1.0$. These results show that, the MZMs at criticality is more robust with increase in the strength of $\gamma$ because the topological phase $w=1$ becomes stable with higher values of $\gamma$. 
  \subsection{Behavior of Multicritical point}\label{topo1}
In the topological systems, multicritical point separates more than two topological phases. Considering the example of Hermitian extended Kitaev chain, it has three $w=0$, $w=1$ and $w=2$ topological phases and two multicritical points at $\lambda_1 = 0, \lambda_2 = 1$ and $\lambda_1 = 2, \lambda_2 = -1$. Authors of the reference \cite{kumar2021multi} has explored the distinct properties of both the multicritical points and have shown that one of the them also acts as the transition point for topological phase transition on the critical line. Features of these two multicritical points in the Hermitian case reveals that, the excitation dispersion at $\lambda_1 = 0, \lambda_2 = 1$ is linear at the gap closing points and it does not break Lorentz invariance. The other multicritical point, $(\lambda_1 = 2, \lambda_2 = -1)$ possess quadratic spectra and it breaks the Lorentz invariance \cite{kumar2021multi,rufo2019multicritical}. 
\begin{figure}[H]
	\centering
	\includegraphics[scale=0.19]{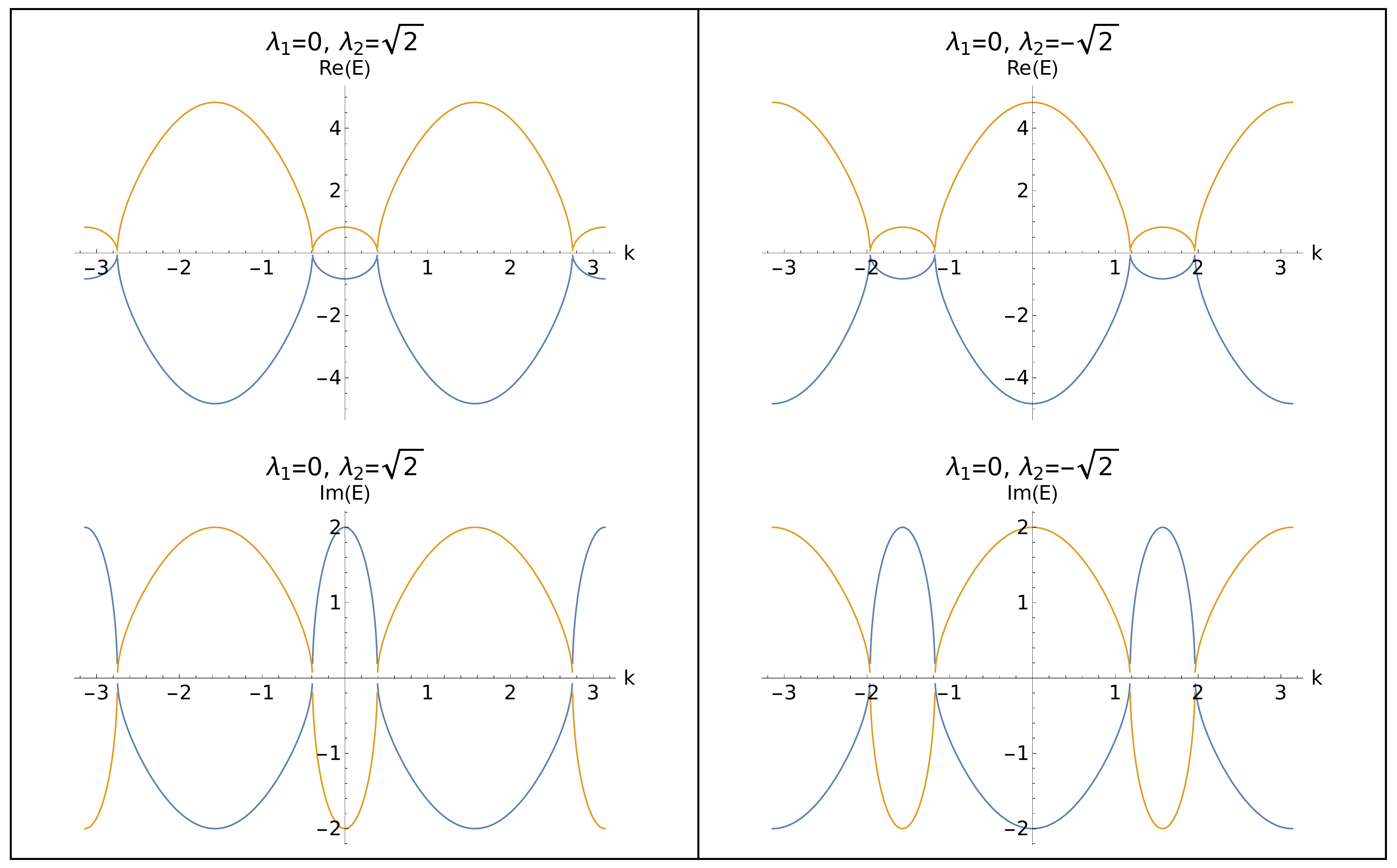}
	\caption{Excitation spectra plotted at both multicritical points. Left and right panel corresponds to the spectrum plotted at two multicritical point locations, $\lambda_1 = 0$, $\lambda_2 = \sqrt{2}$ (q) and $\lambda_1 = 0$, $\lambda_2 = -\sqrt{2}$ (r) where $\mu=1$ and $\gamma=1$. }
	\label{pd2}
\end{figure}
With the spirit of investigating the multicritical points in the non-Hermitian counterpart shows an interesting deviation from the results of Hermitian case. In the non-Hermitian case, multicritical points exists at, $\lambda_1 = 0$, $\lambda_2 = \sqrt{\mu^2 + \gamma^2}$ (q) and $\lambda_1 = 0$, $\lambda_2 = -\sqrt{\mu^2 + \gamma^2}$ (r) (For the points "q" and "r", refer the phase diagram Fig.\ref{pd}). Since the excitation spectra is complex, both real and imaginary dispersion is investigated to understand the nature of the dispersion. The nature of the complex dispersion at the multicritical points "q" and "r" is presented in the Fig.\ref{pd2}. The gap closing points at these multicritical points appear in the vicinity of $k=0, \pm \pi, \pm \pi/2$ which shows the gap closing behavior at criticality. For different critical lines, gap closing occurs at different points which is yet to be explored for different topological phase transitions. From the Fig.\ref{pd2}, it can be seen that, the dispersion remains linear at the gap closing points in both left and right panel which are plotted for two multicritical points respectively. An interesting observation from this analysis is that, the multicritical point which possessed quadratic spectra at the gap closing points in the Hermitian case, now possess linear spectra at the gap closing points in the non-Hermitian configuration. This indicates the change in the nature of multicritical point in the present study.
\subsection{Effect of non-Hermiticity on the topological phases}\label{topo}
\begin{figure}[h]
	\includegraphics[scale=0.21]{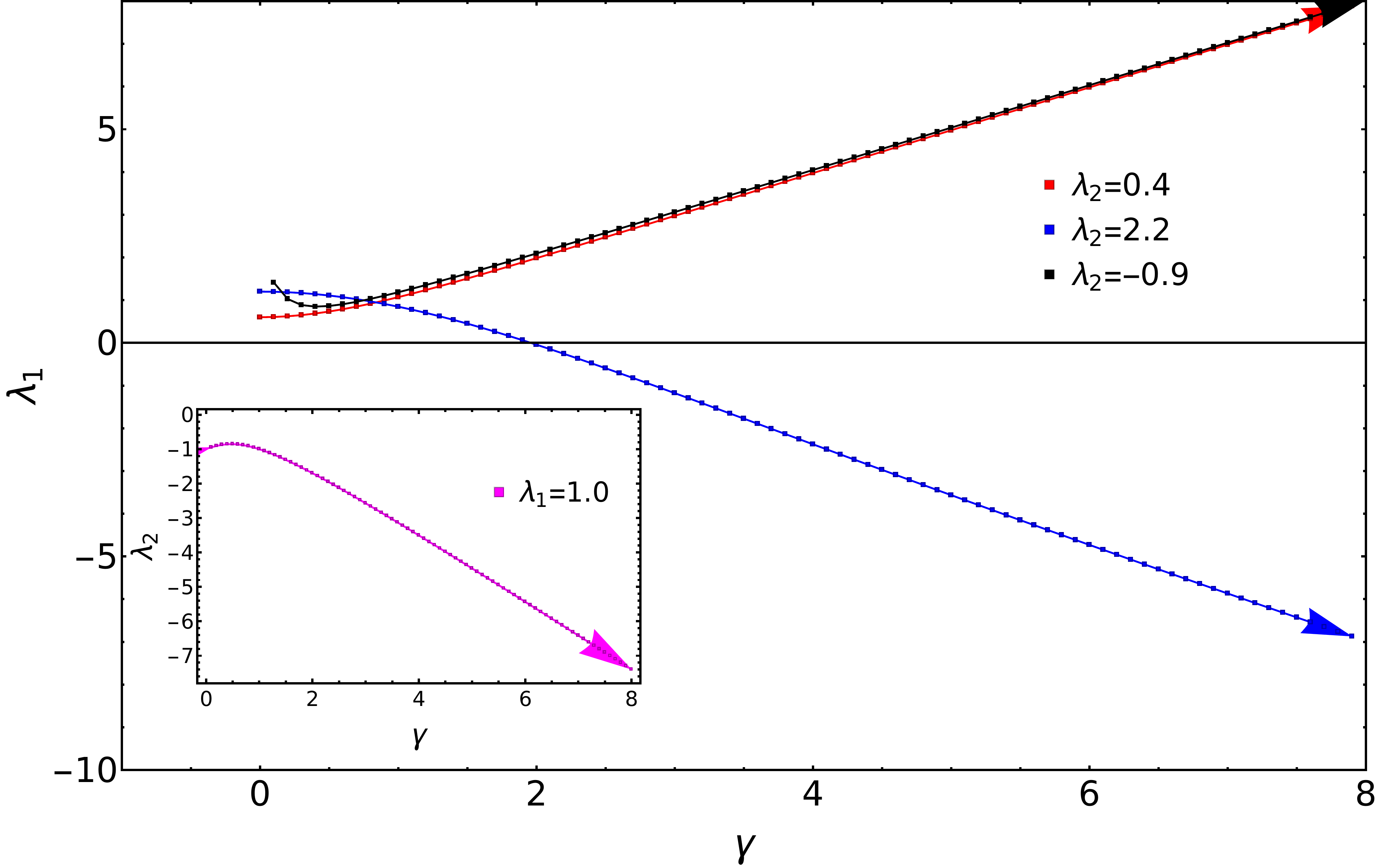}
	\caption{Movement of critical lines plotted with respect to $\gamma$ for fixed values of $\lambda_1$ and $\lambda_2$. }
	\label{fi}
\end{figure}

In the Hermitian case, extended Kitaev chain has three critical lines and supports three gapped phases. The additional feature of the Hermitian extended Kitaev chain is the presence of multicritical point at $\lambda_1 = 2$ and $\lambda_2 = -1$ which has been extensively studied in reference.\cite{rahul2021majorana,kumar2021multi}.
From the ZMS the new critical lines were identified and the phase diagram is constructed (Fig.\ref{pd}). As presented in the Fig.\ref{pass}, the positions of the critical lines change for different values of $\gamma$. Considering topological phase transition between all the topological phases, the critical points move as shown in the Fig.\ref{fi}. Fig.\ref{fi} presents the movement of critical points with respect to $\gamma$ for fixed values of $\lambda_1$ and $\lambda_2$. 
Red, black and blue curves of Fig.\ref{fi} corresponds to $\lambda_2 = 0.4, 2.2$ and $-0.9$ representing topological phase transitions $w = 0$ to $1$, $w=1$ to $2$ and $w=0$ to $1$. An interesting behavior is seen for $w=0$ to $1$ transition in the negative $\lambda_2$ region where the critical points does not continously increase with respect to $\gamma$ like the red curve, instead they decrease initally and then increase with respect to the increase in the value of $\gamma$. This happens because, critical point for $\lambda_2 = -0.9$ move backwards first and then moves forward towards higher values of $\lambda_1$. For $w=0$ to $1$ and $w=2$ to $1$ topological phase transitions, this does not occur since critical points at respective $\lambda_2$ values move forward towards higher values of $\lambda_1$. The arrow mark shows the direction of the movement of the critical points. Blue curve corresponding to $w=2$ to $1$ decreases and moves eventually to the negative $\lambda_1$ plain because, as the value of $\gamma$ is increased, the critical point for $\lambda_2 = 2.2$ moves backward and eventually goes to negative $\lambda_1$ plain because the same critical line is continued in the negative $\lambda_1$ plain. Both red and blue curves show increase and decrease respectively unlike the black curve.\\
Inset of the Fig.\ref{fi} corresponds to the critical points plotted with respect to $\gamma$ for fixed value of $\lambda_1 = 1.0$. This represent the transition from $w=1$ to $2$ topological phases. Even in this phase transition, there is a slight increase in the values of critical points and as the value of $\gamma$ is increased, the value of the critical points decrease continously. This is also because of the backward movement for certain value of $\gamma$ and the forward movement for the rest of the values of $\gamma$.  
	\begin{figure}[h] 
		\includegraphics[scale=0.22]{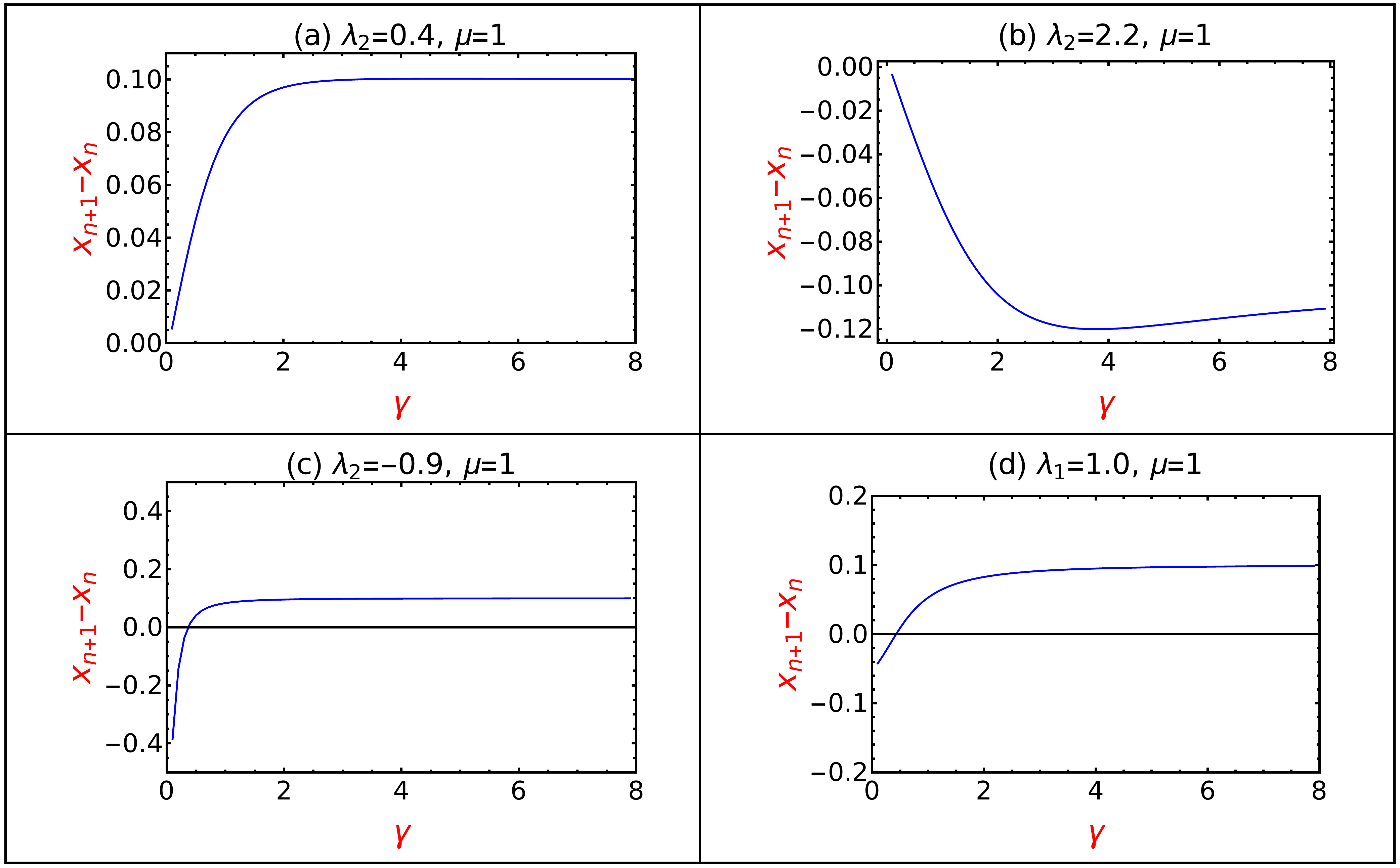}
		\caption{Difference of critical points $(x_{n+1} - x_{n})$ is plotted with respect to $\gamma$ for fixed values of $\lambda_1$ and $\lambda_2$ values.}
		\label{if}
	\end{figure}
Studying the movement and the direction of the critical points with respect to $\gamma$ for all the transitions, the rate of receding in both positive and negative direction with respect to $\gamma$ is also studied which is presented in the Fig.\ref{if}. 
The critical point for corresponding value of $\gamma$ is taken into account starting and the difference of critical points $(x_{n+1} - x_{n})$ with its previous value is calculated and this difference is plotted with respect to the $\gamma$. The quantity $x_{n+1} - x_{n}$ signifying the rate of receding of critical points reveals some interesting behavior in the Fig.\ref{if}. Plots (a) and (b) in the upper panel corresponds to the values of $\lambda_2 = 0.4$ and $2.2$. These values represent the transition from $w=0$ to $1$ and $w=2$ to $1$ topological phases. In the plot.a, the difference, $x_{n+1} - x_{n}$ rises sharply and saturates at 1 signifying that spacing between the critical points increases rapidly till $\gamma = 2$ and for $\gamma> 2$, the spacing remains constant which is dipicted by the saturation of the curve. The quantity, $x_{n+1} - x_{n}$ is negative in the plot.b signifying only the backward movement of critical points with respect to $\gamma$ and it does not mean that spacing between the critical points are negative. Plot.b corresponds to $w=2$ to $1$ topological phase transition and as the plot shows, the difference becomes large with increasing $\gamma$ value but in the opposite direction. Critical points show an interesting behavior in the negative quadrant of $\lambda_2$ (plot.c and d) where they show both and forward movement with respect to increase in the values of $\gamma$. Due to the modification of critical lines where they are no longer linear and also with he displacement of multicritical point, they show the corresponding behavior (refer Fig.\ref{if}) for increasing values of $\gamma$. This analysis shows a detailed effects of $\gamma$ on hte topological phases and criticality.  
\section{Conclusion}
In this work we have studied the topological phases, criticalities, MZMs at criticality and  nature of multicritical points in the non-Hermitian extended Kitaev chain. We have observed a significant changes in the model Hamiltonian and its topological properties with the introduction of non-Hermitian factor $\gamma$. We have computed the critical lines using the method of ZMS since the conventional method of finding the critical lines becomes tedious. With the critical lines obtained from ZMS, we observe and investigate an interesting phenomenon where the critical lines get modified and also it shows the disappearance of multicritical point from it its usual location. Parametric curve study has been performed in order to validate the nature of the critical lines. We also have investigated the MZMs at criticality and its stability with changing values of $\gamma$. We study the behavior of the modified multicritical point and show that the unlike in the Hermitian case, these two multicritical points acquire a similar nature. Finally we study the effect of $\gamma$ on the topological phases by looking at the rate at which the critical points recede. Critical points are calculated for corresponding values of $\gamma$ and its difference is plotted which gives the rate of receding which gives the dynamics of critical lines for increasing values of $\gamma$.   
	\section{Acknowledgments} 
The authors would like to acknowledge DST (EMR/2017/000898) for the funding and RRI library for the books and journals. Finally authors would like to acknowledge ICTS Lectures/seminars/workshops/conferences/discussion meetings of different aspects in physics.
\bibliography{ref.bib}

\begin{thebibliography}{23}%
\makeatletter
\providecommand \@ifxundefined [1]{%
 \@ifx{#1\undefined}
}%
\providecommand \@ifnum [1]{%
 \ifnum #1\expandafter \@firstoftwo
 \else \expandafter \@secondoftwo
 \fi
}%
\providecommand \@ifx [1]{%
 \ifx #1\expandafter \@firstoftwo
 \else \expandafter \@secondoftwo
 \fi
}%
\providecommand \natexlab [1]{#1}%
\providecommand \enquote  [1]{``#1''}%
\providecommand \bibnamefont  [1]{#1}%
\providecommand \bibfnamefont [1]{#1}%
\providecommand \citenamefont [1]{#1}%
\providecommand \href@noop [0]{\@secondoftwo}%
\providecommand \href [0]{\begingroup \@sanitize@url \@href}%
\providecommand \@href[1]{\@@startlink{#1}\@@href}%
\providecommand \@@href[1]{\endgroup#1\@@endlink}%
\providecommand \@sanitize@url [0]{\catcode `\\12\catcode `\$12\catcode
  `\&12\catcode `\#12\catcode `\^12\catcode `\_12\catcode `\%12\relax}%
\providecommand \@@startlink[1]{}%
\providecommand \@@endlink[0]{}%
\providecommand \url  [0]{\begingroup\@sanitize@url \@url }%
\providecommand \@url [1]{\endgroup\@href {#1}{\urlprefix }}%
\providecommand \urlprefix  [0]{URL }%
\providecommand \Eprint [0]{\href }%
\providecommand \doibase [0]{http://dx.doi.org/}%
\providecommand \selectlanguage [0]{\@gobble}%
\providecommand \bibinfo  [0]{\@secondoftwo}%
\providecommand \bibfield  [0]{\@secondoftwo}%
\providecommand \translation [1]{[#1]}%
\providecommand \BibitemOpen [0]{}%
\providecommand \bibitemStop [0]{}%
\providecommand \bibitemNoStop [0]{.\EOS\space}%
\providecommand \EOS [0]{\spacefactor3000\relax}%
\providecommand \BibitemShut  [1]{\csname bibitem#1\endcsname}%
\let\auto@bib@innerbib\@empty
\bibitem [{\citenamefont {Shen}\ \emph {et~al.}(2018)\citenamefont {Shen},
  \citenamefont {Zhen},\ and\ \citenamefont {Fu}}]{shen2018topological}%
  \BibitemOpen
  \bibfield  {author} {\bibinfo {author} {\bibfnamefont {H.}~\bibnamefont
  {Shen}}, \bibinfo {author} {\bibfnamefont {B.}~\bibnamefont {Zhen}}, \ and\
  \bibinfo {author} {\bibfnamefont {L.}~\bibnamefont {Fu}},\ }\href@noop {}
  {\bibfield  {journal} {\bibinfo  {journal} {Physical review letters}\
  }\textbf {\bibinfo {volume} {120}},\ \bibinfo {pages} {146402} (\bibinfo
  {year} {2018})}\BibitemShut {NoStop}%
\bibitem [{\citenamefont {Bergholtz}\ \emph {et~al.}(2019)\citenamefont
  {Bergholtz}, \citenamefont {Budich},\ and\ \citenamefont
  {Kunst}}]{bergholtz2019exceptional}%
  \BibitemOpen
  \bibfield  {author} {\bibinfo {author} {\bibfnamefont {E.~J.}\ \bibnamefont
  {Bergholtz}}, \bibinfo {author} {\bibfnamefont {J.~C.}\ \bibnamefont
  {Budich}}, \ and\ \bibinfo {author} {\bibfnamefont {F.~K.}\ \bibnamefont
  {Kunst}},\ }\href@noop {} {\bibfield  {journal} {\bibinfo  {journal} {arXiv
  preprint arXiv:1912.10048}\ } (\bibinfo {year} {2019})}\BibitemShut {NoStop}%
\bibitem [{\citenamefont {Kunst}\ \emph {et~al.}(2018)\citenamefont {Kunst},
  \citenamefont {Edvardsson}, \citenamefont {Budich},\ and\ \citenamefont
  {Bergholtz}}]{kunst2018biorthogonal}%
  \BibitemOpen
  \bibfield  {author} {\bibinfo {author} {\bibfnamefont {F.~K.}\ \bibnamefont
  {Kunst}}, \bibinfo {author} {\bibfnamefont {E.}~\bibnamefont {Edvardsson}},
  \bibinfo {author} {\bibfnamefont {J.~C.}\ \bibnamefont {Budich}}, \ and\
  \bibinfo {author} {\bibfnamefont {E.~J.}\ \bibnamefont {Bergholtz}},\
  }\href@noop {} {\bibfield  {journal} {\bibinfo  {journal} {Physical review
  letters}\ }\textbf {\bibinfo {volume} {121}},\ \bibinfo {pages} {026808}
  (\bibinfo {year} {2018})}\BibitemShut {NoStop}%
\bibitem [{\citenamefont {R{\"u}ter}\ \emph {et~al.}(2010)\citenamefont
  {R{\"u}ter}, \citenamefont {Makris}, \citenamefont {El-Ganainy},
  \citenamefont {Christodoulides}, \citenamefont {Segev},\ and\ \citenamefont
  {Kip}}]{ruter2010observation}%
  \BibitemOpen
  \bibfield  {author} {\bibinfo {author} {\bibfnamefont {C.~E.}\ \bibnamefont
  {R{\"u}ter}}, \bibinfo {author} {\bibfnamefont {K.~G.}\ \bibnamefont
  {Makris}}, \bibinfo {author} {\bibfnamefont {R.}~\bibnamefont {El-Ganainy}},
  \bibinfo {author} {\bibfnamefont {D.~N.}\ \bibnamefont {Christodoulides}},
  \bibinfo {author} {\bibfnamefont {M.}~\bibnamefont {Segev}}, \ and\ \bibinfo
  {author} {\bibfnamefont {D.}~\bibnamefont {Kip}},\ }\href@noop {} {\bibfield
  {journal} {\bibinfo  {journal} {Nature physics}\ }\textbf {\bibinfo {volume}
  {6}},\ \bibinfo {pages} {192} (\bibinfo {year} {2010})}\BibitemShut {NoStop}%
\bibitem [{\citenamefont {Peng}\ \emph {et~al.}(2014)\citenamefont {Peng},
  \citenamefont {{\"O}zdemir}, \citenamefont {Lei}, \citenamefont {Monifi},
  \citenamefont {Gianfreda}, \citenamefont {Long}, \citenamefont {Fan},
  \citenamefont {Nori}, \citenamefont {Bender},\ and\ \citenamefont
  {Yang}}]{peng2014parity}%
  \BibitemOpen
  \bibfield  {author} {\bibinfo {author} {\bibfnamefont {B.}~\bibnamefont
  {Peng}}, \bibinfo {author} {\bibfnamefont {{\c{S}}.~K.}\ \bibnamefont
  {{\"O}zdemir}}, \bibinfo {author} {\bibfnamefont {F.}~\bibnamefont {Lei}},
  \bibinfo {author} {\bibfnamefont {F.}~\bibnamefont {Monifi}}, \bibinfo
  {author} {\bibfnamefont {M.}~\bibnamefont {Gianfreda}}, \bibinfo {author}
  {\bibfnamefont {G.~L.}\ \bibnamefont {Long}}, \bibinfo {author}
  {\bibfnamefont {S.}~\bibnamefont {Fan}}, \bibinfo {author} {\bibfnamefont
  {F.}~\bibnamefont {Nori}}, \bibinfo {author} {\bibfnamefont {C.~M.}\
  \bibnamefont {Bender}}, \ and\ \bibinfo {author} {\bibfnamefont
  {L.}~\bibnamefont {Yang}},\ }\href@noop {} {\bibfield  {journal} {\bibinfo
  {journal} {Nature Physics}\ }\textbf {\bibinfo {volume} {10}},\ \bibinfo
  {pages} {394} (\bibinfo {year} {2014})}\BibitemShut {NoStop}%
\bibitem [{\citenamefont {Xiao}\ \emph {et~al.}(2017)\citenamefont {Xiao},
  \citenamefont {Zhan}, \citenamefont {Bian}, \citenamefont {Wang},
  \citenamefont {Zhang}, \citenamefont {Wang}, \citenamefont {Li},
  \citenamefont {Mochizuki}, \citenamefont {Kim}, \citenamefont {Kawakami}
  \emph {et~al.}}]{xiao2017observation}%
  \BibitemOpen
  \bibfield  {author} {\bibinfo {author} {\bibfnamefont {L.}~\bibnamefont
  {Xiao}}, \bibinfo {author} {\bibfnamefont {X.}~\bibnamefont {Zhan}}, \bibinfo
  {author} {\bibfnamefont {Z.}~\bibnamefont {Bian}}, \bibinfo {author}
  {\bibfnamefont {K.}~\bibnamefont {Wang}}, \bibinfo {author} {\bibfnamefont
  {X.}~\bibnamefont {Zhang}}, \bibinfo {author} {\bibfnamefont
  {X.}~\bibnamefont {Wang}}, \bibinfo {author} {\bibfnamefont {J.}~\bibnamefont
  {Li}}, \bibinfo {author} {\bibfnamefont {K.}~\bibnamefont {Mochizuki}},
  \bibinfo {author} {\bibfnamefont {D.}~\bibnamefont {Kim}}, \bibinfo {author}
  {\bibfnamefont {N.}~\bibnamefont {Kawakami}},  \emph {et~al.},\ }\href@noop
  {} {\bibfield  {journal} {\bibinfo  {journal} {Nature Physics}\ }\textbf
  {\bibinfo {volume} {13}},\ \bibinfo {pages} {1117} (\bibinfo {year}
  {2017})}\BibitemShut {NoStop}%
\bibitem [{\citenamefont {Ashida}\ \emph {et~al.}(2020)\citenamefont {Ashida},
  \citenamefont {Gong},\ and\ \citenamefont {Ueda}}]{ashida2020non}%
  \BibitemOpen
  \bibfield  {author} {\bibinfo {author} {\bibfnamefont {Y.}~\bibnamefont
  {Ashida}}, \bibinfo {author} {\bibfnamefont {Z.}~\bibnamefont {Gong}}, \ and\
  \bibinfo {author} {\bibfnamefont {M.}~\bibnamefont {Ueda}},\ }\href@noop {}
  {\bibfield  {journal} {\bibinfo  {journal} {Advances in Physics}\ }\textbf
  {\bibinfo {volume} {69}},\ \bibinfo {pages} {249} (\bibinfo {year}
  {2020})}\BibitemShut {NoStop}%
\bibitem [{\citenamefont {Cai}(2021)}]{cai2021boundary}%
  \BibitemOpen
  \bibfield  {author} {\bibinfo {author} {\bibfnamefont {X.}~\bibnamefont
  {Cai}},\ }\href@noop {} {\bibfield  {journal} {\bibinfo  {journal} {Physical
  Review B}\ }\textbf {\bibinfo {volume} {103}},\ \bibinfo {pages} {014201}
  (\bibinfo {year} {2021})}\BibitemShut {NoStop}%
\bibitem [{\citenamefont {El-Ganainy}\ \emph {et~al.}(2019)\citenamefont
  {El-Ganainy}, \citenamefont {Khajavikhan}, \citenamefont {Christodoulides},\
  and\ \citenamefont {Ozdemir}}]{el2019dawn}%
  \BibitemOpen
  \bibfield  {author} {\bibinfo {author} {\bibfnamefont {R.}~\bibnamefont
  {El-Ganainy}}, \bibinfo {author} {\bibfnamefont {M.}~\bibnamefont
  {Khajavikhan}}, \bibinfo {author} {\bibfnamefont {D.~N.}\ \bibnamefont
  {Christodoulides}}, \ and\ \bibinfo {author} {\bibfnamefont {S.~K.}\
  \bibnamefont {Ozdemir}},\ }\href@noop {} {\bibfield  {journal} {\bibinfo
  {journal} {Communications Physics}\ }\textbf {\bibinfo {volume} {2}},\
  \bibinfo {pages} {1} (\bibinfo {year} {2019})}\BibitemShut {NoStop}%
\bibitem [{\citenamefont {El-Ganainy}\ \emph {et~al.}(2018)\citenamefont
  {El-Ganainy}, \citenamefont {Makris}, \citenamefont {Khajavikhan},
  \citenamefont {Musslimani}, \citenamefont {Rotter},\ and\ \citenamefont
  {Christodoulides}}]{el2018non}%
  \BibitemOpen
  \bibfield  {author} {\bibinfo {author} {\bibfnamefont {R.}~\bibnamefont
  {El-Ganainy}}, \bibinfo {author} {\bibfnamefont {K.~G.}\ \bibnamefont
  {Makris}}, \bibinfo {author} {\bibfnamefont {M.}~\bibnamefont {Khajavikhan}},
  \bibinfo {author} {\bibfnamefont {Z.~H.}\ \bibnamefont {Musslimani}},
  \bibinfo {author} {\bibfnamefont {S.}~\bibnamefont {Rotter}}, \ and\ \bibinfo
  {author} {\bibfnamefont {D.~N.}\ \bibnamefont {Christodoulides}},\
  }\href@noop {} {\bibfield  {journal} {\bibinfo  {journal} {Nature Physics}\
  }\textbf {\bibinfo {volume} {14}},\ \bibinfo {pages} {11} (\bibinfo {year}
  {2018})}\BibitemShut {NoStop}%
\bibitem [{\citenamefont {Wang}\ \emph {et~al.}(2015)\citenamefont {Wang},
  \citenamefont {Liu}, \citenamefont {Xiong},\ and\ \citenamefont
  {Tong}}]{wang2015spontaneous}%
  \BibitemOpen
  \bibfield  {author} {\bibinfo {author} {\bibfnamefont {X.}~\bibnamefont
  {Wang}}, \bibinfo {author} {\bibfnamefont {T.}~\bibnamefont {Liu}}, \bibinfo
  {author} {\bibfnamefont {Y.}~\bibnamefont {Xiong}}, \ and\ \bibinfo {author}
  {\bibfnamefont {P.}~\bibnamefont {Tong}},\ }\href@noop {} {\bibfield
  {journal} {\bibinfo  {journal} {Physical Review A}\ }\textbf {\bibinfo
  {volume} {92}},\ \bibinfo {pages} {012116} (\bibinfo {year}
  {2015})}\BibitemShut {NoStop}%
\bibitem [{\citenamefont {Zeng}\ \emph {et~al.}(2016)\citenamefont {Zeng},
  \citenamefont {Zhu}, \citenamefont {Chen}, \citenamefont {You},\ and\
  \citenamefont {L{\"u}}}]{zeng2016non}%
  \BibitemOpen
  \bibfield  {author} {\bibinfo {author} {\bibfnamefont {Q.-B.}\ \bibnamefont
  {Zeng}}, \bibinfo {author} {\bibfnamefont {B.}~\bibnamefont {Zhu}}, \bibinfo
  {author} {\bibfnamefont {S.}~\bibnamefont {Chen}}, \bibinfo {author}
  {\bibfnamefont {L.}~\bibnamefont {You}}, \ and\ \bibinfo {author}
  {\bibfnamefont {R.}~\bibnamefont {L{\"u}}},\ }\href@noop {} {\bibfield
  {journal} {\bibinfo  {journal} {Physical Review A}\ }\textbf {\bibinfo
  {volume} {94}},\ \bibinfo {pages} {022119} (\bibinfo {year}
  {2016})}\BibitemShut {NoStop}%
\bibitem [{\citenamefont {Li}\ \emph {et~al.}(2018)\citenamefont {Li},
  \citenamefont {Zhang}, \citenamefont {Zhang},\ and\ \citenamefont
  {Song}}]{li2018topological}%
  \BibitemOpen
  \bibfield  {author} {\bibinfo {author} {\bibfnamefont {C.}~\bibnamefont
  {Li}}, \bibinfo {author} {\bibfnamefont {X.}~\bibnamefont {Zhang}}, \bibinfo
  {author} {\bibfnamefont {G.}~\bibnamefont {Zhang}}, \ and\ \bibinfo {author}
  {\bibfnamefont {Z.}~\bibnamefont {Song}},\ }\href@noop {} {\bibfield
  {journal} {\bibinfo  {journal} {Physical Review B}\ }\textbf {\bibinfo
  {volume} {97}},\ \bibinfo {pages} {115436} (\bibinfo {year}
  {2018})}\BibitemShut {NoStop}%
\bibitem [{\citenamefont {Lieu}(2018)}]{lieu2018topological}%
  \BibitemOpen
  \bibfield  {author} {\bibinfo {author} {\bibfnamefont {S.}~\bibnamefont
  {Lieu}},\ }\href@noop {} {\bibfield  {journal} {\bibinfo  {journal} {Physical
  Review B}\ }\textbf {\bibinfo {volume} {97}},\ \bibinfo {pages} {045106}
  (\bibinfo {year} {2018})}\BibitemShut {NoStop}%
\bibitem [{\citenamefont {Zhu}\ \emph {et~al.}(2014)\citenamefont {Zhu},
  \citenamefont {L{\"u}},\ and\ \citenamefont {Chen}}]{zhu2014pt}%
  \BibitemOpen
  \bibfield  {author} {\bibinfo {author} {\bibfnamefont {B.}~\bibnamefont
  {Zhu}}, \bibinfo {author} {\bibfnamefont {R.}~\bibnamefont {L{\"u}}}, \ and\
  \bibinfo {author} {\bibfnamefont {S.}~\bibnamefont {Chen}},\ }\href@noop {}
  {\bibfield  {journal} {\bibinfo  {journal} {Physical Review A}\ }\textbf
  {\bibinfo {volume} {89}},\ \bibinfo {pages} {062102} (\bibinfo {year}
  {2014})}\BibitemShut {NoStop}%
\bibitem [{\citenamefont {He}\ and\ \citenamefont {Chien}(2020)}]{he2020non}%
  \BibitemOpen
  \bibfield  {author} {\bibinfo {author} {\bibfnamefont {Y.}~\bibnamefont
  {He}}\ and\ \bibinfo {author} {\bibfnamefont {C.-C.}\ \bibnamefont {Chien}},\
  }\href@noop {} {\bibfield  {journal} {\bibinfo  {journal} {Journal of
  Physics: Condensed Matter}\ }\textbf {\bibinfo {volume} {33}},\ \bibinfo
  {pages} {085501} (\bibinfo {year} {2020})}\BibitemShut {NoStop}%
\bibitem [{\citenamefont {Yao}\ and\ \citenamefont {Wang}(2018)}]{yao2018edge}%
  \BibitemOpen
  \bibfield  {author} {\bibinfo {author} {\bibfnamefont {S.}~\bibnamefont
  {Yao}}\ and\ \bibinfo {author} {\bibfnamefont {Z.}~\bibnamefont {Wang}},\
  }\href@noop {} {\bibfield  {journal} {\bibinfo  {journal} {Physical review
  letters}\ }\textbf {\bibinfo {volume} {121}},\ \bibinfo {pages} {086803}
  (\bibinfo {year} {2018})}\BibitemShut {NoStop}%
\bibitem [{\citenamefont {Yin}\ \emph {et~al.}(2018)\citenamefont {Yin},
  \citenamefont {Jiang}, \citenamefont {Li}, \citenamefont {L{\"u}},\ and\
  \citenamefont {Chen}}]{yin2018geometrical}%
  \BibitemOpen
  \bibfield  {author} {\bibinfo {author} {\bibfnamefont {C.}~\bibnamefont
  {Yin}}, \bibinfo {author} {\bibfnamefont {H.}~\bibnamefont {Jiang}}, \bibinfo
  {author} {\bibfnamefont {L.}~\bibnamefont {Li}}, \bibinfo {author}
  {\bibfnamefont {R.}~\bibnamefont {L{\"u}}}, \ and\ \bibinfo {author}
  {\bibfnamefont {S.}~\bibnamefont {Chen}},\ }\href@noop {} {\bibfield
  {journal} {\bibinfo  {journal} {Physical Review A}\ }\textbf {\bibinfo
  {volume} {97}},\ \bibinfo {pages} {052115} (\bibinfo {year}
  {2018})}\BibitemShut {NoStop}%
\bibitem [{\citenamefont {Navarro-Labastida}\ \emph {et~al.}(2021)\citenamefont
  {Navarro-Labastida}, \citenamefont {Dom{\'\i}nguez-Serna},\ and\
  \citenamefont {Rojas}}]{navarro2021geometrical}%
  \BibitemOpen
  \bibfield  {author} {\bibinfo {author} {\bibfnamefont {L.~A.}\ \bibnamefont
  {Navarro-Labastida}}, \bibinfo {author} {\bibfnamefont {F.~A.}\ \bibnamefont
  {Dom{\'\i}nguez-Serna}}, \ and\ \bibinfo {author} {\bibfnamefont
  {F.}~\bibnamefont {Rojas}},\ }\href@noop {} {\bibfield  {journal} {\bibinfo
  {journal} {arXiv preprint arXiv:2106.02756}\ } (\bibinfo {year}
  {2021})}\BibitemShut {NoStop}%
\bibitem [{\citenamefont {Rahul}\ \emph {et~al.}(2021)\citenamefont {Rahul},
  \citenamefont {Kumar}, \citenamefont {Kartik},\ and\ \citenamefont
  {Sarkar}}]{rahul2021majorana}%
  \BibitemOpen
  \bibfield  {author} {\bibinfo {author} {\bibfnamefont {S.}~\bibnamefont
  {Rahul}}, \bibinfo {author} {\bibfnamefont {R.~R.}\ \bibnamefont {Kumar}},
  \bibinfo {author} {\bibfnamefont {Y.}~\bibnamefont {Kartik}}, \ and\ \bibinfo
  {author} {\bibfnamefont {S.}~\bibnamefont {Sarkar}},\ }\href@noop {}
  {\bibfield  {journal} {\bibinfo  {journal} {Journal of the Physical Society
  of Japan}\ }\textbf {\bibinfo {volume} {90}},\ \bibinfo {pages} {094706}
  (\bibinfo {year} {2021})}\BibitemShut {NoStop}%
\bibitem [{\citenamefont {Verresen}\ \emph {et~al.}(2018)\citenamefont
  {Verresen}, \citenamefont {Jones},\ and\ \citenamefont
  {Pollmann}}]{verresen2018topology}%
  \BibitemOpen
  \bibfield  {author} {\bibinfo {author} {\bibfnamefont {R.}~\bibnamefont
  {Verresen}}, \bibinfo {author} {\bibfnamefont {N.~G.}\ \bibnamefont {Jones}},
  \ and\ \bibinfo {author} {\bibfnamefont {F.}~\bibnamefont {Pollmann}},\
  }\href@noop {} {\bibfield  {journal} {\bibinfo  {journal} {Physical review
  letters}\ }\textbf {\bibinfo {volume} {120}},\ \bibinfo {pages} {057001}
  (\bibinfo {year} {2018})}\BibitemShut {NoStop}%
\bibitem [{\citenamefont {Kumar}\ \emph {et~al.}(2021)\citenamefont {Kumar},
  \citenamefont {Kartik}, \citenamefont {Rahul},\ and\ \citenamefont
  {Sarkar}}]{kumar2021multi}%
  \BibitemOpen
  \bibfield  {author} {\bibinfo {author} {\bibfnamefont {R.~R.}\ \bibnamefont
  {Kumar}}, \bibinfo {author} {\bibfnamefont {Y.}~\bibnamefont {Kartik}},
  \bibinfo {author} {\bibfnamefont {S.}~\bibnamefont {Rahul}}, \ and\ \bibinfo
  {author} {\bibfnamefont {S.}~\bibnamefont {Sarkar}},\ }\href@noop {}
  {\bibfield  {journal} {\bibinfo  {journal} {Scientific Reports}\ }\textbf
  {\bibinfo {volume} {11}},\ \bibinfo {pages} {1} (\bibinfo {year}
  {2021})}\BibitemShut {NoStop}%
\bibitem [{\citenamefont {Rufo}\ \emph {et~al.}(2019)\citenamefont {Rufo},
  \citenamefont {Lopes}, \citenamefont {Continentino},\ and\ \citenamefont
  {Griffith}}]{rufo2019multicritical}%
  \BibitemOpen
  \bibfield  {author} {\bibinfo {author} {\bibfnamefont {S.}~\bibnamefont
  {Rufo}}, \bibinfo {author} {\bibfnamefont {N.}~\bibnamefont {Lopes}},
  \bibinfo {author} {\bibfnamefont {M.~A.}\ \bibnamefont {Continentino}}, \
  and\ \bibinfo {author} {\bibfnamefont {M.}~\bibnamefont {Griffith}},\
  }\href@noop {} {\bibfield  {journal} {\bibinfo  {journal} {Physical Review
  B}\ }\textbf {\bibinfo {volume} {100}},\ \bibinfo {pages} {195432} (\bibinfo
  {year} {2019})}\BibitemShut {NoStop}%
\end{thebibliography}%
\section*{Appendix}
\subsection*{A. Zero mode solutions}
The model Hamiltonian can be written as,
\begin{equation}
H_k = \chi_{z}(k) \sigma_z + \chi_{y}(k) \sigma_y,
\label{1}
\end{equation}
where $ \chi_{z} (k) = 2 \lambda_1 \cos k + 2 \lambda_2 \cos 2k - 2(\mu+ i \gamma),$ and $ \chi_{y} (k) = 2 \lambda_1 \sin k + 2 \lambda_2 \sin 2k.$\\
Substituting the exponential forms of $\cos k$ and $\sin k$, Eq.\ref{1} becomes, 
\begin{multline}
H = \left[ 2 \lambda_1 \frac{1}{2} (e^{-ik} + e^{ik}) + 2 \lambda_2 \frac{1}{2} (e^{-2ik} + e^{2ik} + 2 (\mu+i \gamma))\right]\sigma_z\\ + i \left[ 2 \lambda_1 \frac{1}{2} (e^{ik} - e^{-ik}) + 2 \lambda_2 \frac{1}{2}(e^{2ik} - e^{-2ik}) \right] \sigma_y.
\label{2} 
\end{multline}  
We replace $e^{-ik} = e^{q}$, Eq.\ref{2} becomes, 
\begin{multline}
H = \left[ 2 \lambda_1 \frac{1}{2} (e^{q} + e^{-q}) + 2 \lambda_2 \frac{1}{2} (e^{2q} + e^{-2q}) + 2 (\mu+i \gamma)\right] \sigma_z\\ + i \left[ 2 \lambda_1 \frac{1}{2} (e^{-q} - e^{q}) + 2 \lambda_2 \frac{1}{2}(e^{-2q} - e^{2q}) \right] \sigma_y. 
\label{3} 
\end{multline}  
To find the zero mode solutions, we make $H^2 = 0$. 
By solving the Eq.\ref{3}, we get, 
\begin{equation}
2 \lambda_1 \cosh q + 2 \lambda_2 \cosh 2q - 2(\mu+i \gamma) = \pm  2 \lambda_1 \sinh q + 2 \lambda_2 \sinh 2q.
\label{4} 
\end{equation}
Eq.\ref{4} shows that there will be more than one solution. Considering the Eq.\ref{3} and squaring both sides with $H=0$, we get,  
\begin{multline}
\left(2 \lambda_1 \cosh q + 2 \lambda_2 \cosh 2q - 2(\mu+i \gamma)\right) \\+ i  \left(2 \lambda_1 \sinh q + 2 \lambda_2 \sinh 2q\right) = 0.
\label{5} 
\end{multline} 
Substituting back the exponential forms to the respective terms, we get, 
\begin{multline}
\left[ 2 \lambda_1 \frac{1}{2} (e^{q} + e^{-q}) + 2 \lambda_2 \frac{1}{2} (e^{2q} + e^{-2q} + 2 (\mu+i \gamma))\right]\\ + \left[ 2 \lambda_1 \frac{1}{2} (e^{-q} - e^{q}) + 2 \lambda_2 \frac{1}{2}(e^{-2q} - e^{2q}) \right] = 0 
\label{6}
\end{multline}
Simplifying the Eq.\ref{6}, we end up with a quadratic equation,
\begin{multline}
2 \lambda_1 \frac{1}{2} e^{q}  + 2 \lambda_2 \frac{1}{2} e^{2q} + 2 (\mu+i \gamma) +  2 \lambda_1 \frac{1}{2} e^{q} + 2 \lambda_2 \frac{1}{2} e^{2q}  = 0.
\label{7}
\end{multline}
Simplifying the Eq.\ref{7} to a quadratic form and substituting $e^q = X$,
\begin{equation} \lambda_2 X^2 +  \lambda_1 X  +  (\mu+i \gamma) = 0.
\label{8}
\end{equation}
The roots of this quadratic Equation is given by, 
\begin{equation}
X = \frac{-\lambda_1 \pm \sqrt{\lambda_1^2 + 4 \lambda_2 (\mu+i \gamma)}}{2 \lambda_2} 
\label{9}
\end{equation}
The roots Eq.\ref{9} are the solutions of zero modes. 
\end{document}